\documentclass[secnumarabic,aps,prl,reprint,groupedaddress]{revtex4-1}
\usepackage{verbatim}
\usepackage[english]{babel}
\usepackage{graphicx}
\usepackage{color}
\usepackage{amsmath}

\begin{document}

\title{Spin transfer due to quantum magnetization fluctuations}

\author{Andrei Zholud$^1$}
\author{Ryan Freeman$^1$}
\author{Rongxing Cao$^1$}
\author{Ajit Srivastava$^1$}
\author{Sergei Urazhdin$^1$}

\affiliation{$^1$Department of Physics, Emory University, Atlanta, GA, USA.}

\begin{abstract}
 We utilize a nanoscale magnetic spin-valve structure to demonstrate that current-induced magnetization fluctuations at cryogenic temperatures result predominantly from the quantum fluctuations enhanced by the spin transfer effect. The demonstrated spin transfer due to quantum magnetization fluctuations is distinguished from the previously established current-induced effects by a non-smooth piecewise-linear dependence of the fluctuation intensity on current. It can be driven not only by the directional flows of spin-polarized electrons, but also by their thermal motion and by scattering of unpolarized electrons. This effect is expected to remain non-negligible even at room temperature, and entails a ubiquitous inelastic contribution to spin-polarizing properties of magnetic interfaces.

\end{abstract}

\maketitle
\begin{figure}
	\includegraphics[scale=1]{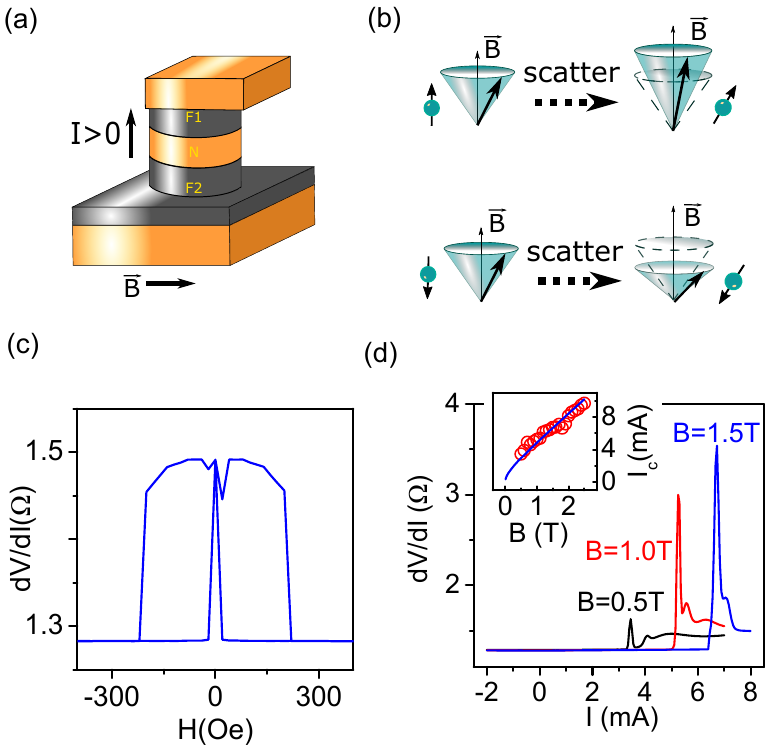}
	\caption{(Color online). (a) Schematic of the tested magnetic nanopillar. (b) Effect of STT on thermal magnetization fluctuations. Top (bottom): scattering of the majority (minority) electron, whose magnetic moment is  parallel (anti-parallel) to the saturating field $\vec{B}$. Spin angular momentum transfer between the electron and the magnetization results in a decrease (increase) of thermal fluctuations. (c) Magnetoelectronic hysteresis loop of the nanopillar. (d) Differential resistance vs current, at the labeled field. The peaks are associated with the onset of dynamical instability at a critical current $I_c$. Inset: $I_c$ vs magnetic field determined from the experimental data (symbols), and the calculation (curve) $I_c=2e\alpha\omega MV/p\mu_B$ based on the Kittel formula for the FMR frequency  $\omega=\gamma\sqrt{B(B+\mu_0M)}$, with Gilbert damping $\alpha=0.02$, polarization $p=0.12$, and the calculated total spin of F1 $MV/\mu_B=1.7\times 10^6$. Here, $\gamma$ is the gyromagnetic ratio, $M$ is the magnetization of Py, $V$ is the volume of F1, and $\mu_B$ is the Bohr magneton. All measurements were performed at temperature $T=3.4$~K.
}\label{fig_1}
\end{figure}

Spin transfer~\cite{Slonczewski1996,Berger1996,Ralph20081190} -- the transfer of angular momentum from spin-polarized electrical current to magnetic materials -- has been extensively researched as an efficient mechanism for the electronic manipulation of the static and dynamic states in nanomagnetic systems, advancing our understanding of nanomagnetism and electronic transport, and enabling the development of energy-efficient magnetic nanodevices~\cite{Tsoi2000,PhysRevLett.84.3149,Kiselev2003,PhysRevLett.92.027201,Ralph20081190,PhysRevB.57.R3213,Lee2004,Tulapurkar2005,Mangin2006,Madami2011,Locatelli2013,Ryu2013,Kent2015}. This effect can be understood based on the argument of spin angular momentum conservation for spin-polarized electrons, scattered by a ferromagnet whose magnetization $\vec{M}$ is not aligned with the direction of polarization. The component of the electron spin transverse to $\vec{M}$ becomes absorbed, exerting a torque on the magnetization termed the spin transfer torque (STT). 
In nanomagnetic devices such as spin valve nanopillars [Fig.~\ref{fig_1}(a)], STT can enhance thermal fluctuations of magnetization [Fig.~\ref{fig_1}(b)], resulting in its reversal~\cite{PhysRevLett.84.3149,Zhang2002}  or auto-oscillation~\cite{Kiselev2003}, which can be utilized in memory, microwave generation, and spin-wave logic~\cite{BehinAein2010,Parkin190}. The approximation for the magnetization as a thermally fluctuating classical vector $\vec{M}$ provides an excellent description for the quasi-uniform magnetization dynamics~\cite{Gurevich1996}. However, the short-wavelength dynamical modes of the magnetization whose frequency extends into the THz range~\cite{White2007} become frozen out at low temperatures, and the effects of spin transfer on them cannot be described in terms of the enhancement or suppression of thermal fluctuations. Short-wavelength modes are not readily accessible to the common electronic spectroscopy and magneto-optical techniques, and their role in spin transfer remains largely unexplored.

Here, we introduce a frequency non-selective, magnetoelectronic measurement approach allowing us to demonstrate that at low temperatures the current-dependent magnetization fluctuations arise predominantly from the enhancement of quantum fluctuations by spin transfer. The observed effect is analogous to the well-studied spontaneous emission of a photon by a two-level system, also caused by quantum fluctuations, which occurs even when there are no photons to stimulate the emission. In the studied magnetic system, the role of photons is played by magnons - the quanta of the dynamical magnetization modes. Our results indicate that the contribution of quantum fluctuations enhanced by spin transfer remains larger than that of thermal fluctuations at temperatures up to over $100$~K, and remains non-negligible even at room temperature. The demonstrated effect also entails a ubiquitous inelastic contribution to spin-polarizing properties of magnetic interfaces.

The effects of STT on thermal magnetization fluctuations [Fig.~\ref{fig_1}(b)] can be described in terms of their current-dependent spectral intensity, or equivalently current-dependent  population of magnons~\cite{Demidov2011}, 
\begin{equation}\label{STT}
<N>=\frac{N_0}{1-I/I_c}
\end{equation}
where $N_0$ is the magnon population in thermal equilibrium, and $I_c$ is the critical current for the onset of the dynamical instability~\cite{Slonczewski1996,Berger1996,Ralph20081190}. 
The dependence Eq.~(\ref{STT}) has been verified by the magneto-optical~\cite{Demidov2011} and magnetoelectronic techniques~\cite{Petit2007}. We utilized the latter to verify the established effects of STT in the Permalloy (Py)-based magnetic nanopillars used in our study. The nanopillars were based on the multilayer with structure Cu(40)Py(10)Cu(4)Py(5)Au(2), where thicknesses are in nanometers. We used a combination of e-beam lithography and Ar ion milling to pattern the "free" layer F1=Py(5) and the Cu(4) spacer into a cylindrical shape with a $70$~nm diameter, while the thicker "polarizer" F2=Py(10) was only partially patterned, allowing the magnons generated in this layer due to spin transfer to escape from the active area. Thus, spin transfer affected only the fluctuations of the free layer F1, while the role of F2 was limited to polarizing the electron current flowing through the nanostructure. The nanopillars were contacted with a Cu(80) top electrode, electrically isolated from the bottom electrode by an insulating SiO$_2$(15) layer. Magnetoelectronic measurements were performed in a pseudo-four probe geometry by the lock-in detection technique, with an ac current of $50$~$\mu$A rms at a frequency of $1.3$~kHz superimposed on the dc bias current.

The dependence of resistance on the magnetic field for our test structure is typical for the giant magnetoresistance (GMR)~\cite{Fert1988} in magnetic nanopillars, Fig.~\ref{fig_1}(c). The current-dependent differential resistance exhibits a sharp peak consistent with the onset of the dynamical instability at the critical current $I_c$~\cite{PhysRevLett.84.3149,Kiselev2003,PhysRevLett.92.027201}, Fig.~\ref{fig_1}(d). The dependence of $I_c$ on the magnetic field agrees with the calculation based on the Kittel formula for the ferromagnetic resonance (FMR) mode~\cite{Kittel}, inset in Fig.~\ref{fig_1}(d).

To introduce our approach to magnetoelectronic measurements of current-dependent magnetization fluctuations, we analyze the relationship between GMR and magnon population. The GMR results in a sinusoidal dependence of resistance $R$ on the angle $\theta$ between the magnetizations of the ``free" layer F1 and the polarizer F2, $R(\theta)=R(0)+\Delta R\sin^2(\theta/2)\approx R(0)+\Delta R\theta^2$, where $R_0$ is the resistance minimum, and $\Delta R$ is the total magnetoresistance~\cite{PhysRevB.67.094421}. The quadratic dependence $R(\theta)$ at small $\theta$ can be viewed as the lowest-order Taylor expansion. By symmetry, quadratic relationship is also expected for non-uniform states, albeit somewhat rescaled by the electron diffusion across magnetically inhomogeneous regions.

To analyze the relation between $\theta$ and the magnon population, we note that each magnon has spin $1$, regardless of the spatial characteristics of the corresponding dynamical mode~\cite{White2007}. Therefore, for the ferromagnet with the total spin $L=MV/\mu_B$, the total magnon population is related to the average $\theta$ by $N=MV\sin^2(\theta/2)/\mu_B$ ~\cite{supplemental}. Here, $V$ is the volume of the nanomagnet, and $\mu_B$ is the Bohr magneton. Thus, resistance is proportional to the total magnon population, $R(\theta)=R(0)+CN\mu_B\Delta R/MV$ with the coefficient $C$ of order $1$ reflecting the contribution of magnetic inhomogeneity to the GMR signal. Therefore, resistance variations due to GMR directly reflect the total magnon population in the nanopillar, not limited to quasi-uniform dynamical modes.

\begin{figure*}
	\includegraphics[scale=1.0]{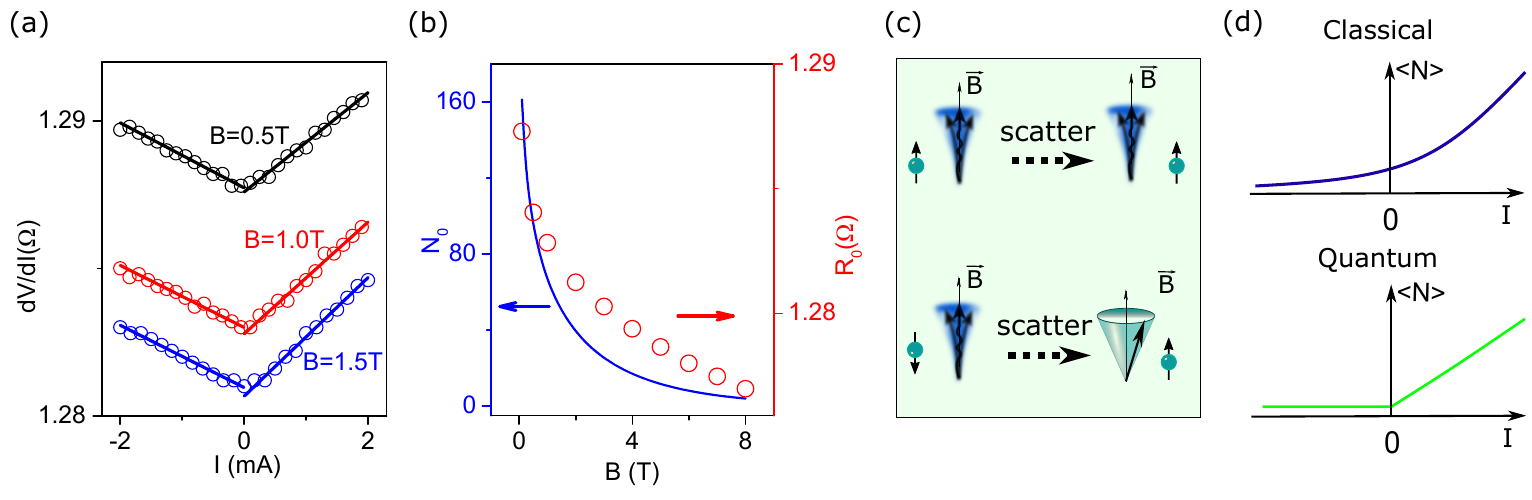}
	\caption{(Color online). (a) Differential resistance vs current, at the labeled values of field and $T=3.4$~K. Symbols: data, lines: best linear fits performed separately for the $I<0$ and $I>0$ data. (b) Differential resistance (symbols, right scale) and the calculated total magnon population (curve, left scale) vs $B$ at $I=0$.	(c)  Schematic of spin transfer due to quantum fluctuations of magnetization.  Top: the magnetic moment of electron is initially aligned with the field (majority electron). The projections of magnetization and of the electron's magnetic moment on the field are maximal, and cannot change in the scattering process. Bottom: the magnetic moment of electron is initially anti-parallel to the field (minority electron). The electron's spin can flip, enhancing fluctuations. (d) Current dependence of magnon population due to spin transfer in the classical limit described by Eq.~(\ref{STT}) (top), and in the quantum limit described by Eq.~(\ref{QST}) at $N_0=0$ (bottom), in the approximation of perfect spin-filtering by the polarizer, $p=1$.} 
	\label{fig_2}
\end{figure*}

At subcritical currents, the resistance of the studied nanopillar exhibits an unusual piecewise-linear dependence, with a weak singularity at $I=0$, and a slope at $I>0$ larger than at $I<0$, Fig.~\ref{fig_2}(a). The variations of the applied field shift the curves, without noticeably affecting their slopes. The shift can be explained by the magnon freeze-out, as illustrated in Fig.~\ref{fig_2}(b) that shows the field dependence of resistance at $I=0$, together with the calculated total thermal magnon population. The calculation was performed in the exchange approximation $\hbar \omega=Dk^2$, using the stiffness $D=4\times10^{-40}$~Jm$^2$ for Permalloy~\cite{PyDispertion}. The magnon population was determined by summing up the contributions $N_0=1/[exp(\hbar\omega/kT)-1]$ of each mode, with the allowed values of wavevector $k$ determined using the pinned-magnetization boundary conditions for a square shape with dimensions $70\times 70\times 5$~nm. The overall agreement between the variations of resistance and the calculated magnon population confirms the relationship between them established above, with the scaling coefficient $C\approx 0.2$ reflecting a reduced sensitivity to short-wavelength modes. The observed dependence $R(B)$ is somewhat weaker than expected based on the calculation of $N_0(B)$, likely because the exchange approximation overestimates the frequencies, and thus underestimates the populations, of short-wavelength modes~\cite{White2007,Gurevich1996}.

Since the field does not noticeably affect the slopes of the curves in Fig.~\ref{fig_2}(a), the observed piecewise-linear dependence cannot be associated with thermal fluctuations whose intensity is controlled by the field, see Fig.~\ref{fig_2}(b). It cannot be explained by Joule heating, because the dissipated power is quadratic in current, so that the increase of resistance due to heating must be also at least quadratic in current. It is also inconsistent with the analytical expression Eq.~(\ref{STT}) of the spin torque theory. Electronic shot noise exhibits a similar linear increase of power with bias~\cite{spietz}. However, shot noise (or fluctuating electron current) can contribute to the measured differential resistance only by inducing magnetization fluctuations, which in the absence of thermal fluctuations is forbidden by the angular momentum conservation argument of spin torque theory.

We conclude that a previously unrecognized contribution to spin transfer, not described as enhancement of thermal magnetization fluctuations, results in a linear in current increase of magnon population. To interpret our observations, we note that even if thermal fluctuations are negligible at low temperature, the spin polarization of electrons scattered by the magnetic system cannot be perfectly aligned with the magnetization because of the quantum fluctuations of the latter, driving electron spin dynamics, and resulting in spin transfer. The proposed quantum effect must be distinct from the established spin torque effects described by Eq.~(\ref{STT}). Indeed, quantum fluctuations cannot be affected by scattering of the majority electrons, since in contrast to thermal fluctuations they cannot be suppressed [Fig.~\ref{fig_2}(c), top]. However, they can be enhanced by scattering of the minority electrons [Fig.~\ref{fig_2}(c), bottom].

There is no established theory for the effects of quantum magnetization fluctuations on spin transfer, although the latter has been analyzed in the context of the quantum theory of magnetism~\cite{PhysRevB.85.092403,PhysRevB.87.174433,wang2013}. Here, we present a simple model that allows us to extend the spin-angular momentum conservation argument underlying the spin torque theory~\cite{Slonczewski1996,Ralph20081190} to quantum magnetization states. We can describe the FMR mode by the dynamical states of a quantum macrospin $\vec{L}$ representing the magnetization~\cite{PhysRevB.69.134430}, whose projection $L_z$ on the z-axis directed opposite to $\vec{B}$ characterizes magnon population $N=L-L_z$. An electron with spin $\vec{s}=(a,b)$ scattered by the magnetic layer experiences exchange interaction $H_{ex}=J_{ex}\vec{s}\cdot\vec{L}/L$, where $J_{ex}$ is the s-d exchange energy. This interaction results in the precession of both $\vec{L}$ and $\vec{s}$ around the total angular momentum $\vec{J}=\vec{L}+\vec{s}$ conserved by the exchange Hamiltonian. This description is a natural extension of the electron spin precession around the magnetization analyzed in the spin torque theory. Following the dephasing argument originally proposed by Slonczewski~\cite{Slonczewski1996}, we can assume that the precession phases are randomized due to variations among electron trajectories. Under these assumptions, one can determine the change of $L_z$, and thus the average number $<\Delta N>$ of magnons generated by the scattered electron. At $N<<L$, we obtain~\cite{PhysRevB.69.134430,supplemental},
\begin{equation}\label{deltaN}	
<\Delta N>=-<\Delta L_z> \approx b^2/L+b^2N/L-a^2N/L.
\end{equation}
 This equation can be interpreted by analogy to the interaction between a two-level system and the electromagnetic field. The two-level system is the spin of the scattered electron, and the role of photons is played by magnons. The first term describes spontaneous emission of magnons, which can occur even in the absence of magnons at $N=0$. The second and the third terms describe stimulated emission and absorption, respectively, with the probability proportional to the number of magnons. This interpretation closely follows the ideas of Berger~\cite{Berger1996}, who described spin transfer in terms of stimulated and spontaneous magnon emission, but ultimately neglected the spontaneous contribution in the analysis of degenerate long-wavelength modes. Without the spontaneous contribution, Eq.~(\ref{deltaN}) is equivalent to the result obtained in the spin torque theory $\frac{\partial\theta}{\partial t}|_{STT}=Ig/eL\sin\theta$, Eq.~(17) in Ref.~\cite{Slonczewski1996}. Here, $g$ is a function of order one determined by the polarization of current $I$. Indeed, using $\Delta n_e=I\Delta t/e$ to represent the number of electrons scattered by the ferromagnet and $N=L(1-\cos\theta)$, we obtain $\frac{\Delta N}{\Delta n_e}=g\sin^2\theta\approx 2gN/L$, consistent with the contribution of stimulated processes in Eq.~(\ref{deltaN}).

In the steady state, the magnon population is determined by the balance between spin transfer driven by current $I$, and the dynamical relaxation. Describing the latter by the Landau damping, or equivalently for small $N$ by the relaxation time approximation $\partial N/\partial t|_{D}=-(N-N_0)/\tau$~\cite{Berger1996,PhysRevB.69.134430} with $\tau=1/(2\alpha\omega)$, we obtain~\cite{supplemental}
\begin{equation}\label{QST}
<N(I)>=\frac{N_0+(|I|/p+I)/2I_c}{1-I/I_c},
\end{equation}
where $p=a^2-b^2$ describes the current polarization. The unusual non-analytical form of Eq.~(\ref{QST}) originates from the asymmetry of Eq.~(\ref{deltaN}) with respect to exchanging $a$ and $b$ describing the current reversal. Equation (\ref{QST}) reduces to the STT result Eq.~(\ref{STT}) in the classical limit, at $N_0\gg1$ [Fig.~\ref{fig_2}(d), top], when the stimulated contribution in Eq.~(\ref{deltaN}) is dominant. In the quantum limit at $N_0\ll 1$, we obtain a piecewise-linear dependence [Fig.~\ref{fig_2}d, bottom]. The data in Fig.~\ref{fig_2}(a) are consistent with the dominant quantum contribution once we account for the imperfect electron spin polarization, $p<1$ in Eq.~(\ref{QST}), resulting in spontaneous magnon generation at both positive and negative currents.

We emphasize that the contribution of quantum fluctuations is negligible for the degenerate quasi-uniform dynamical modes. However,  at $3.4$~K the modes with frequencies above $~300$~GHz are frozen out. Since exchange interaction between the electron spin and the magnetization underlying spin transfer is local, Eq.~(\ref{QST}) with appropriate values of $I_c$ must be also applicable to these modes. A similar argument has been put forward in spin torque theory~\cite{Polianski}.
A direct summation of Eq.~(\ref{QST}) over the entire magnon spectrum confirmed that the quantum contribution to the current-dependent magnon population is dominant at $3.4$~K, consistent with our interpretation of Fig.~\ref{fig_2}(a)~\cite{supplemental}.

\begin{figure}
	\includegraphics[scale=1.0]{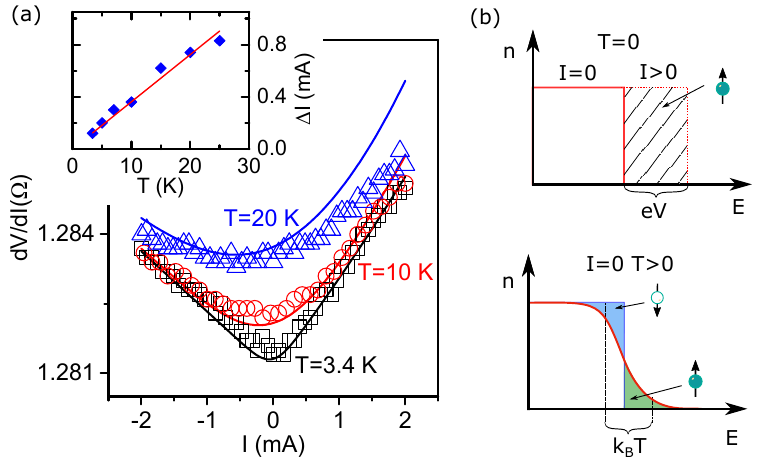}
	\caption{(Color online).
	(a) Symbols: differential resistance vs current, at the labeled values of temperature and $B=1$~T. Curves: calculations based on Eq.~(\ref{QST}) summed over all the magnon modes, and convoluted with a Gaussian of FWHM $\Delta I=1.9kT/eR_0$. The vertical scale was determined from the relation between $N(B)$ and $R_0(B)$ in Fig.~\ref{fig_2}(b). Symbols in the inset show the broadening determined by fitting the data with a piecewise-linear dependence convolved with a Gaussian, the line is $\Delta I=1.9kT/eR_0$. 
	(b) Schematics of thermal broadening effects. Top: at $T=0$, the Fermi distribution of scattered electrons is step-like. Bias current shifts the distribution, driving the spin transfer. Bottom: at finite $T$, scattering of thermally excited electrons and holes occurs even at zero bias, with the effect equivalent to that of a bias distribution with the width $k_BT/e$.
}\label{fig_3}
\end{figure}

The role of quantum fluctuations in spin transfer was further elucidated by measurements at higher temperatures, where we observe a rapid broadening of the zero-current singularity  [Fig.~\ref{fig_3}(a)]. This broadening cannot be attributed to the increasing role of thermal magnetization fluctuations, since the piecewise-linear dependence is still apparent at larger currents even at $20$~K. To analyze this effect, we fit the data with a piecewise-linear dependence convolved with a Gaussian. The extracted broadening width closely follows a linear dependence $\Delta I=(1.9\pm 0.1)kT/eR_0$, inset in Fig.~\ref{fig_3}(a). 

A calculation based on the summation of Eq.~(\ref{QST}) convolved with a Gaussian of width $1.9 kT/eR_0$ [curves Fig.~\ref{fig_3}(a)] reproduces the thermal broadening effect, but somewhat exaggerates the classical contribution at higher temperatures, likely due to the overestimated frequencies, and thus the relaxation rates, of high-frequency magnons in the exchange approximation for magnon dispersion used in our calculation. We note that this calculation used only the parameter values extracted from matching the calculated magnon populations with resistance at $T=3.4$~K, Fig.~\ref{fig_3}(b). A good agreement with the data, achieved at elevated temperatures without any fitting parameters, supports the validity of our model.

The observed thermal broadening is consistent with the proposed quantum mechanism. Bias current shifts the electron distribution in the magnetic nanopillar, driving the spin transfer [Fig.~\ref{fig_3}(b), top]. At finite temperature, the electron distribution becomes thermally broadened, resulting in scattering of thermally excited electrons and holes [Fig.~\ref{fig_3}(b), bottom] equivalent to a distribution of width $\Delta V=kT/e$ of the bias voltage applied to F1, facilitating spin transfer even in the absence of directional current flow. The relation $\Delta I=(1.9\pm 0.1)kT/eR_0$ obtained by fitting the data [inset in Fig.~\ref{fig_3}(a)] is consistent with the approximately equal contributions of layers F1 and F2 to the total resistance $R_0$, such that $\Delta V\approx IR_0/2$.

\begin{figure}
	\includegraphics[scale=1.1]{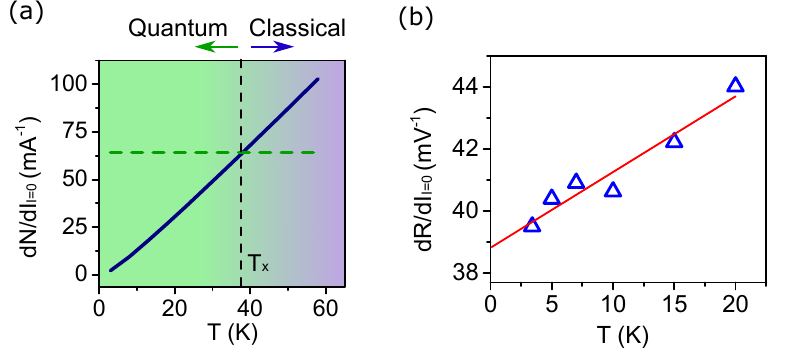}
	\caption{(Color online). (a) The calculated slope $dN/dI$, at $I=0$ and $B=1$~T, for the classical (blue curve) and quantum (horizontal green line) contributions to the dependence of magnon population on current. $T_x$ marks the crossover temperature between the quantum and the classical regimes.
	(b) Symbols: The slope $dR/dI$ of resistance vs current at $I=0$ and $B=1$~T, determined by fitting the data with a piecewise-linear dependence convoluted with a Gaussian of full width $\Delta I=1.9kT/eR_0$. Line: best linear fit of the data, with the intercept of $39$~mV$^{-1}$ and slope $0.245$~mV$^{-1}$K$^{-1}$. The extrapolated quantum-to-classical crossover temperature is $T_x=160$~K. 
	}\label{fig_4}
\end{figure}

Thermal broadening washes out the singular piecewise-linear dependence, but the contribution of quantum fluctuations to spin transfer remains significant even at elevated temperatures. 
Since the slopes of the piecewise-linear dependence are different for positive and negative currents, the value of $dN/dI$ at $I=0$ remains finite even in the presence of thermal broadening. By convolving the dependence $N(R)$ [Eq.~(\ref{QST})] with a Gaussian and differentiating with respect to $I$, we obtain $dN/dI=1/2I_c$ at $I=0$, independent of temperature.
For STT facilitated by thermal fluctuations, the slope is $dN/dI=N_0/I_c$, see Eq.~(\ref{STT}). Since the quantum contribution to spin transfer is independent of $T$ at $I=0$, the corresponding value of $dN/dI$ for the total magnon population is independent of $T$, horizontal line in Fig.~\ref{fig_4}(a). For the STT contribution facilitated by thermal fluctuations, the calculated value increases almost linearly with temperature (blue curve), indicating that this contribution is dominated by the degenerately populated low-frequency modes described by the Rayleigh-Jeans law. At $B=1$~T, the calculated crossover from predominantly quantum to the classical (thermal) spin transfer regime occurs at temperature $T_x=38$~K. The slope $dR/dI$ at $I=0$, determined from our measurements, increases linearly with temperature [Fig.~\ref{fig_4}(b)], in agreement with our model. The $T=0$ intercept represents the quantum contribution, and the slope reflects the classical one. The value $T_x=160$~K extrapolated from these data is larger than the calculated value, likely because the quantum contribution is underestimated in the model based on the parabolic magnon dispersion. Based on our data, we estimate the characteristic frequency of magnons involved in spin transfer, $f_0=k_BT_x/h\approx 3.5$~THz.

Quantum fluctuations can significantly contribute to current-induced phenomena whenever highly nonuniform dynamical states are involved, for example in reversal via fast domain wall motion in technologically important nanomagnets with perpendicular magnetic anisotropy~\cite{Mangin2006}. More generally, the demonstrated magnon generation mechanism can decrease the effective magnetization, lowering the reversal barriers. The relative contribution of quantum fluctuations to current-induced phenomena in antiferromagnets~\cite{Jungwirth2016} is likely larger than in ferromagnets, since the characteristic magnon frequencies are almost two orders of magnitude higher, resulting in significantly smaller thermal magnon populations. Quantum fluctuations may contribute to other phenomena involving interaction between magnetization and conduction electrons, including spin-orbit effects~\cite{Ando2008,Liu555,Nakayama2013}, optically-driven effects~\cite{Turgut2016,NatComm82017,Rudolf2012}, and spin-caloritronic effects~\cite{PhysRevLett.99.066603,PhysRevLett.95.016601,Wegrowe2010,PhysRevB.94.064414,Choi2015}. They may also provide a significant contribution to spin-polarizing properties of ferromagnets. Indeed, according to Eq.~(\ref{deltaN}), the probability for a spin-down conduction electron to spin-flip while generating an FMR magnon at $T=0$ is $1/L$, where $L$ is the total spin of the ferromagnet. Since the total number of magnetic modes is approximately $L$, and each mode contributes to such spin flipping, the total probability of spin-flip is of the order one, providing a ubiquitous inelastic contribution to the spin-polarization of electrons flowing through ferromagnets. This may explain why even advanced spin-dependent band structure calculations generally underestimate electron spin flipping rates at magnetic interfaces~\cite{Belashchenko}. By tailoring the magnon spectrum via material and geometry engineering, it may become possible to control the effects of quantum magnetization fluctuations on magnetoelectronic phenomena in nanomagnetic systems.

We acknowledge support from NSF Grant Nos. ECCS-1509794 and DMR-1504449.

\bibliography{bibl}{}

\begin{thebibliography}{46}%
\makeatletter
\providecommand \@ifxundefined [1]{%
 \@ifx{#1\undefined}
}%
\providecommand \@ifnum [1]{%
 \ifnum #1\expandafter \@firstoftwo
 \else \expandafter \@secondoftwo
 \fi
}%
\providecommand \@ifx [1]{%
 \ifx #1\expandafter \@firstoftwo
 \else \expandafter \@secondoftwo
 \fi
}%
\providecommand \natexlab [1]{#1}%
\providecommand \enquote  [1]{``#1''}%
\providecommand \bibnamefont  [1]{#1}%
\providecommand \bibfnamefont [1]{#1}%
\providecommand \citenamefont [1]{#1}%
\providecommand \href@noop [0]{\@secondoftwo}%
\providecommand \href [0]{\begingroup \@sanitize@url \@href}%
\providecommand \@href[1]{\@@startlink{#1}\@@href}%
\providecommand \@@href[1]{\endgroup#1\@@endlink}%
\providecommand \@sanitize@url [0]{\catcode `\\12\catcode `\$12\catcode
  `\&12\catcode `\#12\catcode `\^12\catcode `\_12\catcode `\%12\relax}%
\providecommand \@@startlink[1]{}%
\providecommand \@@endlink[0]{}%
\providecommand \url  [0]{\begingroup\@sanitize@url \@url }%
\providecommand \@url [1]{\endgroup\@href {#1}{\urlprefix }}%
\providecommand \urlprefix  [0]{URL }%
\providecommand \Eprint [0]{\href }%
\providecommand \doibase [0]{http://dx.doi.org/}%
\providecommand \selectlanguage [0]{\@gobble}%
\providecommand \bibinfo  [0]{\@secondoftwo}%
\providecommand \bibfield  [0]{\@secondoftwo}%
\providecommand \translation [1]{[#1]}%
\providecommand \BibitemOpen [0]{}%
\providecommand \bibitemStop [0]{}%
\providecommand \bibitemNoStop [0]{.\EOS\space}%
\providecommand \EOS [0]{\spacefactor3000\relax}%
\providecommand \BibitemShut  [1]{\csname bibitem#1\endcsname}%
\let\auto@bib@innerbib\@empty
%</preamble>
\bibitem [{\citenamefont {Slonczewski}(1996)}]{Slonczewski1996}%
  \BibitemOpen
  \bibfield  {author} {\bibinfo {author} {\bibfnamefont {J.}~\bibnamefont
  {Slonczewski}},\ }\href {\doibase 10.1016/0304-8853(96)00062-5} {\bibfield
  {journal} {\bibinfo  {journal} {J. Magn. Magn. Mater.}\ }\textbf {\bibinfo
  {volume} {159}},\ \bibinfo {pages} {L1} (\bibinfo {year} {1996})}\BibitemShut
  {NoStop}%
\bibitem [{\citenamefont {Berger}(1996)}]{Berger1996}%
  \BibitemOpen
  \bibfield  {author} {\bibinfo {author} {\bibfnamefont {L.}~\bibnamefont
  {Berger}},\ }\href {http://prb.aps.org/abstract/PRB/v54/i13/p9353{\_}1}
  {\bibfield  {journal} {\bibinfo  {journal} {Phys. Rev. B}\ }\textbf {\bibinfo
  {volume} {54}},\ \bibinfo {pages} {9353} (\bibinfo {year}
  {1996})}\BibitemShut {NoStop}%
\bibitem [{\citenamefont {Ralph}\ and\ \citenamefont
  {Stiles}(2008)}]{Ralph20081190}%
  \BibitemOpen
  \bibfield  {author} {\bibinfo {author} {\bibfnamefont {D.}~\bibnamefont
  {Ralph}}\ and\ \bibinfo {author} {\bibfnamefont {M.}~\bibnamefont {Stiles}},\
  }\href {\doibase http://dx.doi.org/10.1016/j.jmmm.2007.12.019} {\bibfield
  {journal} {\bibinfo  {journal} {Journal of Magnetism and Magnetic Materials}\
  }\textbf {\bibinfo {volume} {320}},\ \bibinfo {pages} {1190 } (\bibinfo
  {year} {2008})}\BibitemShut {NoStop}%
\bibitem [{\citenamefont {Tsoi}\ \emph {et~al.}(2000)\citenamefont {Tsoi},
  \citenamefont {Jansen}, \citenamefont {Bass}, \citenamefont {Chiang},
  \citenamefont {Tsoi},\ and\ \citenamefont {Wyder}}]{Tsoi2000}%
  \BibitemOpen
  \bibfield  {author} {\bibinfo {author} {\bibfnamefont {M.}~\bibnamefont
  {Tsoi}}, \bibinfo {author} {\bibfnamefont {A.~G.~M.}\ \bibnamefont {Jansen}},
  \bibinfo {author} {\bibfnamefont {J.}~\bibnamefont {Bass}}, \bibinfo {author}
  {\bibfnamefont {W.-C.}\ \bibnamefont {Chiang}}, \bibinfo {author}
  {\bibfnamefont {V.}~\bibnamefont {Tsoi}}, \ and\ \bibinfo {author}
  {\bibfnamefont {P.}~\bibnamefont {Wyder}},\ }\href {\doibase
  10.1038/35017512} {\bibfield  {journal} {\bibinfo  {journal} {Nature}\
  }\textbf {\bibinfo {volume} {406}},\ \bibinfo {pages} {46} (\bibinfo {year}
  {2000})}\BibitemShut {NoStop}%
\bibitem [{\citenamefont {Katine}\ \emph {et~al.}(2000)\citenamefont {Katine},
  \citenamefont {Albert}, \citenamefont {Buhrman}, \citenamefont {Myers},\ and\
  \citenamefont {Ralph}}]{PhysRevLett.84.3149}%
  \BibitemOpen
  \bibfield  {author} {\bibinfo {author} {\bibfnamefont {J.~A.}\ \bibnamefont
  {Katine}}, \bibinfo {author} {\bibfnamefont {F.~J.}\ \bibnamefont {Albert}},
  \bibinfo {author} {\bibfnamefont {R.~A.}\ \bibnamefont {Buhrman}}, \bibinfo
  {author} {\bibfnamefont {E.~B.}\ \bibnamefont {Myers}}, \ and\ \bibinfo
  {author} {\bibfnamefont {D.~C.}\ \bibnamefont {Ralph}},\ }\href {\doibase
  10.1103/PhysRevLett.84.3149} {\bibfield  {journal} {\bibinfo  {journal}
  {Phys. Rev. Lett.}\ }\textbf {\bibinfo {volume} {84}},\ \bibinfo {pages}
  {3149} (\bibinfo {year} {2000})}\BibitemShut {NoStop}%
\bibitem [{\citenamefont {Kiselev}\ \emph {et~al.}(2003)\citenamefont
  {Kiselev}, \citenamefont {Sankey}, \citenamefont {Krivorotov}, \citenamefont
  {Emley}, \citenamefont {Schoelkopf}, \citenamefont {Buhrman},\ and\
  \citenamefont {Ralph}}]{Kiselev2003}%
  \BibitemOpen
  \bibfield  {author} {\bibinfo {author} {\bibfnamefont {S.~I.}\ \bibnamefont
  {Kiselev}}, \bibinfo {author} {\bibfnamefont {J.~C.}\ \bibnamefont {Sankey}},
  \bibinfo {author} {\bibfnamefont {I.~N.}\ \bibnamefont {Krivorotov}},
  \bibinfo {author} {\bibfnamefont {N.~C.}\ \bibnamefont {Emley}}, \bibinfo
  {author} {\bibfnamefont {R.~J.}\ \bibnamefont {Schoelkopf}}, \bibinfo
  {author} {\bibfnamefont {R.~A.}\ \bibnamefont {Buhrman}}, \ and\ \bibinfo
  {author} {\bibfnamefont {D.~C.}\ \bibnamefont {Ralph}},\ }\href {\doibase
  10.1038/nature01967} {\bibfield  {journal} {\bibinfo  {journal} {Nature}\
  }\textbf {\bibinfo {volume} {425}},\ \bibinfo {pages} {380} (\bibinfo {year}
  {2003})}\BibitemShut {NoStop}%
\bibitem [{\citenamefont {Rippard}\ \emph {et~al.}(2004)\citenamefont
  {Rippard}, \citenamefont {Pufall}, \citenamefont {Kaka}, \citenamefont
  {Russek},\ and\ \citenamefont {Silva}}]{PhysRevLett.92.027201}%
  \BibitemOpen
  \bibfield  {author} {\bibinfo {author} {\bibfnamefont {W.~H.}\ \bibnamefont
  {Rippard}}, \bibinfo {author} {\bibfnamefont {M.~R.}\ \bibnamefont {Pufall}},
  \bibinfo {author} {\bibfnamefont {S.}~\bibnamefont {Kaka}}, \bibinfo {author}
  {\bibfnamefont {S.~E.}\ \bibnamefont {Russek}}, \ and\ \bibinfo {author}
  {\bibfnamefont {T.~J.}\ \bibnamefont {Silva}},\ }\href {\doibase
  10.1103/PhysRevLett.92.027201} {\bibfield  {journal} {\bibinfo  {journal}
  {Phys. Rev. Lett.}\ }\textbf {\bibinfo {volume} {92}},\ \bibinfo {pages}
  {027201} (\bibinfo {year} {2004})}\BibitemShut {NoStop}%
\bibitem [{\citenamefont {Bazaliy}\ \emph {et~al.}(1998)\citenamefont
  {Bazaliy}, \citenamefont {Jones},\ and\ \citenamefont
  {Zhang}}]{PhysRevB.57.R3213}%
  \BibitemOpen
  \bibfield  {author} {\bibinfo {author} {\bibfnamefont {Y.~B.}\ \bibnamefont
  {Bazaliy}}, \bibinfo {author} {\bibfnamefont {B.~A.}\ \bibnamefont {Jones}},
  \ and\ \bibinfo {author} {\bibfnamefont {S.-C.}\ \bibnamefont {Zhang}},\
  }\href {\doibase 10.1103/PhysRevB.57.R3213} {\bibfield  {journal} {\bibinfo
  {journal} {Phys. Rev. B}\ }\textbf {\bibinfo {volume} {57}},\ \bibinfo
  {pages} {R3213} (\bibinfo {year} {1998})}\BibitemShut {NoStop}%
\bibitem [{\citenamefont {Lee}\ \emph {et~al.}(2004)\citenamefont {Lee},
  \citenamefont {Deac}, \citenamefont {Redon}, \citenamefont {Nozi{\`{e}}res},\
  and\ \citenamefont {Dieny}}]{Lee2004}%
  \BibitemOpen
  \bibfield  {author} {\bibinfo {author} {\bibfnamefont {K.-J.}\ \bibnamefont
  {Lee}}, \bibinfo {author} {\bibfnamefont {A.}~\bibnamefont {Deac}}, \bibinfo
  {author} {\bibfnamefont {O.}~\bibnamefont {Redon}}, \bibinfo {author}
  {\bibfnamefont {J.-P.}\ \bibnamefont {Nozi{\`{e}}res}}, \ and\ \bibinfo
  {author} {\bibfnamefont {B.}~\bibnamefont {Dieny}},\ }\href {\doibase
  10.1038/nmat1237} {\bibfield  {journal} {\bibinfo  {journal} {Nature
  Materials}\ }\textbf {\bibinfo {volume} {3}},\ \bibinfo {pages} {877}
  (\bibinfo {year} {2004})}\BibitemShut {NoStop}%
\bibitem [{\citenamefont {Tulapurkar}\ \emph {et~al.}(2005)\citenamefont
  {Tulapurkar}, \citenamefont {Suzuki}, \citenamefont {Fukushima},
  \citenamefont {Kubota}, \citenamefont {Maehara}, \citenamefont {Tsunekawa},
  \citenamefont {Djayaprawira}, \citenamefont {Watanabe},\ and\ \citenamefont
  {Yuasa}}]{Tulapurkar2005}%
  \BibitemOpen
  \bibfield  {author} {\bibinfo {author} {\bibfnamefont {A.~A.}\ \bibnamefont
  {Tulapurkar}}, \bibinfo {author} {\bibfnamefont {Y.}~\bibnamefont {Suzuki}},
  \bibinfo {author} {\bibfnamefont {A.}~\bibnamefont {Fukushima}}, \bibinfo
  {author} {\bibfnamefont {H.}~\bibnamefont {Kubota}}, \bibinfo {author}
  {\bibfnamefont {H.}~\bibnamefont {Maehara}}, \bibinfo {author} {\bibfnamefont
  {K.}~\bibnamefont {Tsunekawa}}, \bibinfo {author} {\bibfnamefont {D.~D.}\
  \bibnamefont {Djayaprawira}}, \bibinfo {author} {\bibfnamefont
  {N.}~\bibnamefont {Watanabe}}, \ and\ \bibinfo {author} {\bibfnamefont
  {S.}~\bibnamefont {Yuasa}},\ }\href {\doibase 10.1038/nature04207} {\bibfield
   {journal} {\bibinfo  {journal} {Nature}\ }\textbf {\bibinfo {volume}
  {438}},\ \bibinfo {pages} {339} (\bibinfo {year} {2005})}\BibitemShut
  {NoStop}%
\bibitem [{\citenamefont {Mangin}\ \emph {et~al.}(2006)\citenamefont {Mangin},
  \citenamefont {Ravelosona}, \citenamefont {Katine}, \citenamefont {Carey},
  \citenamefont {Terris},\ and\ \citenamefont {Fullerton}}]{Mangin2006}%
  \BibitemOpen
  \bibfield  {author} {\bibinfo {author} {\bibfnamefont {S.}~\bibnamefont
  {Mangin}}, \bibinfo {author} {\bibfnamefont {D.}~\bibnamefont {Ravelosona}},
  \bibinfo {author} {\bibfnamefont {J.~A.}\ \bibnamefont {Katine}}, \bibinfo
  {author} {\bibfnamefont {M.~J.}\ \bibnamefont {Carey}}, \bibinfo {author}
  {\bibfnamefont {B.~D.}\ \bibnamefont {Terris}}, \ and\ \bibinfo {author}
  {\bibfnamefont {E.~E.}\ \bibnamefont {Fullerton}},\ }\href {\doibase
  10.1038/nmat1595} {\bibfield  {journal} {\bibinfo  {journal} {Nature
  Materials}\ }\textbf {\bibinfo {volume} {5}},\ \bibinfo {pages} {210}
  (\bibinfo {year} {2006})}\BibitemShut {NoStop}%
\bibitem [{\citenamefont {Madami}\ \emph {et~al.}(2011)\citenamefont {Madami},
  \citenamefont {Bonetti}, \citenamefont {Consolo}, \citenamefont {Tacchi},
  \citenamefont {Carlotti}, \citenamefont {Gubbiotti}, \citenamefont {Mancoff},
  \citenamefont {Yar},\ and\ \citenamefont {{\AA}kerman}}]{Madami2011}%
  \BibitemOpen
  \bibfield  {author} {\bibinfo {author} {\bibfnamefont {M.}~\bibnamefont
  {Madami}}, \bibinfo {author} {\bibfnamefont {S.}~\bibnamefont {Bonetti}},
  \bibinfo {author} {\bibfnamefont {G.}~\bibnamefont {Consolo}}, \bibinfo
  {author} {\bibfnamefont {S.}~\bibnamefont {Tacchi}}, \bibinfo {author}
  {\bibfnamefont {G.}~\bibnamefont {Carlotti}}, \bibinfo {author}
  {\bibfnamefont {G.}~\bibnamefont {Gubbiotti}}, \bibinfo {author}
  {\bibfnamefont {F.~B.}\ \bibnamefont {Mancoff}}, \bibinfo {author}
  {\bibfnamefont {M.~A.}\ \bibnamefont {Yar}}, \ and\ \bibinfo {author}
  {\bibfnamefont {J.}~\bibnamefont {{\AA}kerman}},\ }\href {\doibase
  10.1038/nnano.2011.140} {\bibfield  {journal} {\bibinfo  {journal} {Nature
  Nanotechnology}\ }\textbf {\bibinfo {volume} {6}},\ \bibinfo {pages} {635}
  (\bibinfo {year} {2011})}\BibitemShut {NoStop}%
\bibitem [{\citenamefont {Locatelli}\ \emph {et~al.}(2013)\citenamefont
  {Locatelli}, \citenamefont {Cros},\ and\ \citenamefont
  {Grollier}}]{Locatelli2013}%
  \BibitemOpen
  \bibfield  {author} {\bibinfo {author} {\bibfnamefont {N.}~\bibnamefont
  {Locatelli}}, \bibinfo {author} {\bibfnamefont {V.}~\bibnamefont {Cros}}, \
  and\ \bibinfo {author} {\bibfnamefont {J.}~\bibnamefont {Grollier}},\ }\href
  {\doibase 10.1038/nmat3823} {\bibfield  {journal} {\bibinfo  {journal}
  {Nature Materials}\ }\textbf {\bibinfo {volume} {13}},\ \bibinfo {pages} {11}
  (\bibinfo {year} {2013})}\BibitemShut {NoStop}%
\bibitem [{\citenamefont {Ryu}\ \emph {et~al.}(2013)\citenamefont {Ryu},
  \citenamefont {Thomas}, \citenamefont {Yang},\ and\ \citenamefont
  {Parkin}}]{Ryu2013}%
  \BibitemOpen
  \bibfield  {author} {\bibinfo {author} {\bibfnamefont {K.-S.}\ \bibnamefont
  {Ryu}}, \bibinfo {author} {\bibfnamefont {L.}~\bibnamefont {Thomas}},
  \bibinfo {author} {\bibfnamefont {S.-H.}\ \bibnamefont {Yang}}, \ and\
  \bibinfo {author} {\bibfnamefont {S.}~\bibnamefont {Parkin}},\ }\href
  {\doibase 10.1038/nnano.2013.102} {\bibfield  {journal} {\bibinfo  {journal}
  {Nature Nanotechnology}\ }\textbf {\bibinfo {volume} {8}},\ \bibinfo {pages}
  {527} (\bibinfo {year} {2013})}\BibitemShut {NoStop}%
\bibitem [{\citenamefont {Kent}\ and\ \citenamefont
  {Worledge}(2015)}]{Kent2015}%
  \BibitemOpen
  \bibfield  {author} {\bibinfo {author} {\bibfnamefont {A.~D.}\ \bibnamefont
  {Kent}}\ and\ \bibinfo {author} {\bibfnamefont {D.~C.}\ \bibnamefont
  {Worledge}},\ }\href {http://dx.doi.org/10.1038/nnano.2015.24} {\bibfield
  {journal} {\bibinfo  {journal} {Nat Nano}\ }\textbf {\bibinfo {volume}
  {10}},\ \bibinfo {pages} {187} (\bibinfo {year} {2015})},\ \bibinfo {note}
  {commentary}\BibitemShut {NoStop}%
\bibitem [{\citenamefont {Zhang}\ \emph {et~al.}(2002)\citenamefont {Zhang},
  \citenamefont {Levy},\ and\ \citenamefont {Fert}}]{Zhang2002}%
  \BibitemOpen
  \bibfield  {author} {\bibinfo {author} {\bibfnamefont {S.}~\bibnamefont
  {Zhang}}, \bibinfo {author} {\bibfnamefont {P.~M.}\ \bibnamefont {Levy}}, \
  and\ \bibinfo {author} {\bibfnamefont {A.}~\bibnamefont {Fert}},\ }\href
  {\doibase 10.1103/physrevlett.88.236601} {\bibfield  {journal} {\bibinfo
  {journal} {Physical Review Letters}\ }\textbf {\bibinfo {volume} {88}}
  (\bibinfo {year} {2002}),\ 10.1103/physrevlett.88.236601}\BibitemShut
  {NoStop}%
\bibitem [{\citenamefont {Behin-Aein}\ \emph {et~al.}(2010)\citenamefont
  {Behin-Aein}, \citenamefont {Datta}, \citenamefont {Salahuddin},\ and\
  \citenamefont {Datta}}]{BehinAein2010}%
  \BibitemOpen
  \bibfield  {author} {\bibinfo {author} {\bibfnamefont {B.}~\bibnamefont
  {Behin-Aein}}, \bibinfo {author} {\bibfnamefont {D.}~\bibnamefont {Datta}},
  \bibinfo {author} {\bibfnamefont {S.}~\bibnamefont {Salahuddin}}, \ and\
  \bibinfo {author} {\bibfnamefont {S.}~\bibnamefont {Datta}},\ }\href
  {\doibase 10.1038/nnano.2010.31} {\bibfield  {journal} {\bibinfo  {journal}
  {Nature Nanotechnology}\ }\textbf {\bibinfo {volume} {5}},\ \bibinfo {pages}
  {266} (\bibinfo {year} {2010})}\BibitemShut {NoStop}%
\bibitem [{\citenamefont {Parkin}\ \emph {et~al.}(2008)\citenamefont {Parkin},
  \citenamefont {Hayashi},\ and\ \citenamefont {Thomas}}]{Parkin190}%
  \BibitemOpen
  \bibfield  {author} {\bibinfo {author} {\bibfnamefont {S.~S.~P.}\
  \bibnamefont {Parkin}}, \bibinfo {author} {\bibfnamefont {M.}~\bibnamefont
  {Hayashi}}, \ and\ \bibinfo {author} {\bibfnamefont {L.}~\bibnamefont
  {Thomas}},\ }\href {\doibase 10.1126/science.1145799} {\bibfield  {journal}
  {\bibinfo  {journal} {Science}\ }\textbf {\bibinfo {volume} {320}},\ \bibinfo
  {pages} {190} (\bibinfo {year} {2008})},\ \Eprint
  {http://arxiv.org/abs/http://science.sciencemag.org/content/320/5873/190.full.pdf}
  {http://science.sciencemag.org/content/320/5873/190.full.pdf} \BibitemShut
  {NoStop}%
\bibitem [{\citenamefont {Gurevich}\ and\ \citenamefont
  {Melkov}(1996)}]{Gurevich1996}%
  \BibitemOpen
  \bibfield  {author} {\bibinfo {author} {\bibfnamefont {A.~G.}\ \bibnamefont
  {Gurevich}}\ and\ \bibinfo {author} {\bibfnamefont {G.~A.}\ \bibnamefont
  {Melkov}},\ }\href
  {https://www.amazon.com/Magnetization-Oscillations-Waves-Alexander-Gurevich/dp/0849394600%3FSubscriptionId%3D0JYN1NVW651KCA56C102%26tag%3Dtechkie-20%26linkCode%3Dxm2%26camp%3D2025%26creative%3D165953%26creativeASIN%3D0849394600}
  {\emph {\bibinfo {title} {Magnetization Oscillations and Waves}}}\ (\bibinfo
  {publisher} {CRC Press},\ \bibinfo {year} {1996})\BibitemShut {NoStop}%
\bibitem [{\citenamefont {White}(2007)}]{White2007}%
  \BibitemOpen
  \bibfield  {author} {\bibinfo {author} {\bibfnamefont {R.~M.}\ \bibnamefont
  {White}},\ }\href {\doibase 10.1007/978-3-540-69025-2} {\emph {\bibinfo
  {title} {Quantum Theory of Magnetism}}}\ (\bibinfo  {publisher} {Springer
  Berlin Heidelberg},\ \bibinfo {year} {2007})\BibitemShut {NoStop}%
\bibitem [{\citenamefont {Demidov}\ \emph {et~al.}(2011)\citenamefont
  {Demidov}, \citenamefont {Urazhdin}, \citenamefont {Edwards}, \citenamefont
  {Stiles}, \citenamefont {McMichael},\ and\ \citenamefont
  {Demokritov}}]{Demidov2011}%
  \BibitemOpen
  \bibfield  {author} {\bibinfo {author} {\bibfnamefont {V.~E.}\ \bibnamefont
  {Demidov}}, \bibinfo {author} {\bibfnamefont {S.}~\bibnamefont {Urazhdin}},
  \bibinfo {author} {\bibfnamefont {E.~R.~J.}\ \bibnamefont {Edwards}},
  \bibinfo {author} {\bibfnamefont {M.~D.}\ \bibnamefont {Stiles}}, \bibinfo
  {author} {\bibfnamefont {R.~D.}\ \bibnamefont {McMichael}}, \ and\ \bibinfo
  {author} {\bibfnamefont {S.~O.}\ \bibnamefont {Demokritov}},\ }\href
  {\doibase 10.1103/physrevlett.107.107204} {\bibfield  {journal} {\bibinfo
  {journal} {Physical Review Letters}\ }\textbf {\bibinfo {volume} {107}}
  (\bibinfo {year} {2011}),\ 10.1103/physrevlett.107.107204}\BibitemShut
  {NoStop}%
\bibitem [{\citenamefont {Petit}\ \emph {et~al.}(2007)\citenamefont {Petit},
  \citenamefont {Baraduc}, \citenamefont {Thirion}, \citenamefont {Ebels},
  \citenamefont {Liu}, \citenamefont {Li}, \citenamefont {Wang},\ and\
  \citenamefont {Dieny}}]{Petit2007}%
  \BibitemOpen
  \bibfield  {author} {\bibinfo {author} {\bibfnamefont {S.}~\bibnamefont
  {Petit}}, \bibinfo {author} {\bibfnamefont {C.}~\bibnamefont {Baraduc}},
  \bibinfo {author} {\bibfnamefont {C.}~\bibnamefont {Thirion}}, \bibinfo
  {author} {\bibfnamefont {U.}~\bibnamefont {Ebels}}, \bibinfo {author}
  {\bibfnamefont {Y.}~\bibnamefont {Liu}}, \bibinfo {author} {\bibfnamefont
  {M.}~\bibnamefont {Li}}, \bibinfo {author} {\bibfnamefont {P.}~\bibnamefont
  {Wang}}, \ and\ \bibinfo {author} {\bibfnamefont {B.}~\bibnamefont {Dieny}},\
  }\href {\doibase 10.1103/physrevlett.98.077203} {\bibfield  {journal}
  {\bibinfo  {journal} {Physical Review Letters}\ }\textbf {\bibinfo {volume}
  {98}} (\bibinfo {year} {2007}),\ 10.1103/physrevlett.98.077203}\BibitemShut
  {NoStop}%
\bibitem [{\citenamefont {Baibich}\ \emph {et~al.}(1988)\citenamefont
  {Baibich}, \citenamefont {Broto}, \citenamefont {Fert}, \citenamefont
  {Van~Dau}, \citenamefont {Petroff}, \citenamefont {Etienne}, \citenamefont
  {Creuzet}, \citenamefont {Friederich},\ and\ \citenamefont
  {Chazelas}}]{Fert1988}%
  \BibitemOpen
  \bibfield  {author} {\bibinfo {author} {\bibfnamefont {M.~N.}\ \bibnamefont
  {Baibich}}, \bibinfo {author} {\bibfnamefont {J.~M.}\ \bibnamefont {Broto}},
  \bibinfo {author} {\bibfnamefont {A.}~\bibnamefont {Fert}}, \bibinfo {author}
  {\bibfnamefont {F.~N.}\ \bibnamefont {Van~Dau}}, \bibinfo {author}
  {\bibfnamefont {F.}~\bibnamefont {Petroff}}, \bibinfo {author} {\bibfnamefont
  {P.}~\bibnamefont {Etienne}}, \bibinfo {author} {\bibfnamefont
  {G.}~\bibnamefont {Creuzet}}, \bibinfo {author} {\bibfnamefont
  {A.}~\bibnamefont {Friederich}}, \ and\ \bibinfo {author} {\bibfnamefont
  {J.}~\bibnamefont {Chazelas}},\ }\href {\doibase 10.1103/PhysRevLett.61.2472}
  {\bibfield  {journal} {\bibinfo  {journal} {Phys. Rev. Lett.}\ }\textbf
  {\bibinfo {volume} {61}},\ \bibinfo {pages} {2472} (\bibinfo {year}
  {1988})}\BibitemShut {NoStop}%
\bibitem [{\citenamefont {Kittel}(1948)}]{Kittel}%
  \BibitemOpen
  \bibfield  {author} {\bibinfo {author} {\bibfnamefont {C.}~\bibnamefont
  {Kittel}},\ }\href {\doibase 10.1103/PhysRev.73.155} {\bibfield  {journal}
  {\bibinfo  {journal} {Phys. Rev.}\ }\textbf {\bibinfo {volume} {73}},\
  \bibinfo {pages} {155} (\bibinfo {year} {1948})}\BibitemShut {NoStop}%
\bibitem [{\citenamefont {Bauer}\ \emph {et~al.}(2003)\citenamefont {Bauer},
  \citenamefont {Tserkovnyak}, \citenamefont {Huertas-Hernando},\ and\
  \citenamefont {Brataas}}]{PhysRevB.67.094421}%
  \BibitemOpen
  \bibfield  {author} {\bibinfo {author} {\bibfnamefont {G.~E.~W.}\
  \bibnamefont {Bauer}}, \bibinfo {author} {\bibfnamefont {Y.}~\bibnamefont
  {Tserkovnyak}}, \bibinfo {author} {\bibfnamefont {D.}~\bibnamefont
  {Huertas-Hernando}}, \ and\ \bibinfo {author} {\bibfnamefont
  {A.}~\bibnamefont {Brataas}},\ }\href {\doibase 10.1103/PhysRevB.67.094421}
  {\bibfield  {journal} {\bibinfo  {journal} {Phys. Rev. B}\ }\textbf {\bibinfo
  {volume} {67}},\ \bibinfo {pages} {094421} (\bibinfo {year}
  {2003})}\BibitemShut {NoStop}%
\bibitem [{sup()}]{supplemental}%
  \BibitemOpen
  \href@noop {} {}\bibinfo {note} {See Supplemental Material for a detailed
  description of the proposed spin-scattering model, relation between magnon
  population and resistance, additional data and simulation of heating
  effects}\BibitemShut {NoStop}%
\bibitem [{\citenamefont {Menzinger}\ \emph {et~al.}(1968)\citenamefont
  {Menzinger}, \citenamefont {Caglioti}, \citenamefont {Shirane}, \citenamefont
  {Nathans}, \citenamefont {Pickart},\ and\ \citenamefont
  {Alperin}}]{PyDispertion}%
  \BibitemOpen
  \bibfield  {author} {\bibinfo {author} {\bibfnamefont {F.}~\bibnamefont
  {Menzinger}}, \bibinfo {author} {\bibfnamefont {G.}~\bibnamefont {Caglioti}},
  \bibinfo {author} {\bibfnamefont {G.}~\bibnamefont {Shirane}}, \bibinfo
  {author} {\bibfnamefont {R.}~\bibnamefont {Nathans}}, \bibinfo {author}
  {\bibfnamefont {S.~J.}\ \bibnamefont {Pickart}}, \ and\ \bibinfo {author}
  {\bibfnamefont {H.~A.}\ \bibnamefont {Alperin}},\ }\href@noop {} {\bibfield
  {journal} {\bibinfo  {journal} {J. Appl. Phys.}\ }\textbf {\bibinfo {volume}
  {39}},\ \bibinfo {pages} {455} (\bibinfo {year} {1968})}\BibitemShut
  {NoStop}%
\bibitem [{\citenamefont {Spietz}\ \emph {et~al.}(2003)\citenamefont {Spietz},
  \citenamefont {Lehnert}, \citenamefont {Siddiqi},\ and\ \citenamefont
  {Schoelkopf}}]{spietz}%
  \BibitemOpen
  \bibfield  {author} {\bibinfo {author} {\bibfnamefont {L.}~\bibnamefont
  {Spietz}}, \bibinfo {author} {\bibfnamefont {K.~W.}\ \bibnamefont {Lehnert}},
  \bibinfo {author} {\bibfnamefont {I.}~\bibnamefont {Siddiqi}}, \ and\
  \bibinfo {author} {\bibfnamefont {R.~J.}\ \bibnamefont {Schoelkopf}},\ }\href
  {\doibase 10.1126/science.1084647} {\bibfield  {journal} {\bibinfo  {journal}
  {Science}\ }\textbf {\bibinfo {volume} {300}},\ \bibinfo {pages} {1929}
  (\bibinfo {year} {2003})}\BibitemShut {NoStop}%
\bibitem [{\citenamefont {Wang}\ and\ \citenamefont
  {Sham}(2012)}]{PhysRevB.85.092403}%
  \BibitemOpen
  \bibfield  {author} {\bibinfo {author} {\bibfnamefont {Y.}~\bibnamefont
  {Wang}}\ and\ \bibinfo {author} {\bibfnamefont {L.~J.}\ \bibnamefont
  {Sham}},\ }\href {\doibase 10.1103/PhysRevB.85.092403} {\bibfield  {journal}
  {\bibinfo  {journal} {Phys. Rev. B}\ }\textbf {\bibinfo {volume} {85}},\
  \bibinfo {pages} {092403} (\bibinfo {year} {2012})}\BibitemShut {NoStop}%
\bibitem [{\citenamefont {Wang}\ and\ \citenamefont
  {Sham}(2013)}]{PhysRevB.87.174433}%
  \BibitemOpen
  \bibfield  {author} {\bibinfo {author} {\bibfnamefont {Y.}~\bibnamefont
  {Wang}}\ and\ \bibinfo {author} {\bibfnamefont {L.~J.}\ \bibnamefont
  {Sham}},\ }\href {\doibase 10.1103/PhysRevB.87.174433} {\bibfield  {journal}
  {\bibinfo  {journal} {Phys. Rev. B}\ }\textbf {\bibinfo {volume} {87}},\
  \bibinfo {pages} {174433} (\bibinfo {year} {2013})}\BibitemShut {NoStop}%
\bibitem [{\citenamefont {Wang}\ \emph {et~al.}(2013)\citenamefont {Wang},
  \citenamefont {Zhou},\ and\ \citenamefont {Zhang}}]{wang2013}%
  \BibitemOpen
  \bibfield  {author} {\bibinfo {author} {\bibfnamefont {Y.}~\bibnamefont
  {Wang}}, \bibinfo {author} {\bibfnamefont {Y.}~\bibnamefont {Zhou}}, \ and\
  \bibinfo {author} {\bibfnamefont {F.-C.}\ \bibnamefont {Zhang}},\ }\href
  {\doibase 10.1063/1.4813320} {\bibfield  {journal} {\bibinfo  {journal}
  {Applied Physics Letters}\ }\textbf {\bibinfo {volume} {103}},\ \bibinfo
  {pages} {022403} (\bibinfo {year} {2013})}\BibitemShut {NoStop}%
\bibitem [{\citenamefont {Urazhdin}(2004)}]{PhysRevB.69.134430}%
  \BibitemOpen
  \bibfield  {author} {\bibinfo {author} {\bibfnamefont {S.}~\bibnamefont
  {Urazhdin}},\ }\href {\doibase 10.1103/PhysRevB.69.134430} {\bibfield
  {journal} {\bibinfo  {journal} {Phys. Rev. B}\ }\textbf {\bibinfo {volume}
  {69}},\ \bibinfo {pages} {134430} (\bibinfo {year} {2004})}\BibitemShut
  {NoStop}%
\bibitem [{\citenamefont {Polianski}\ and\ \citenamefont
  {Brouwer}(2004)}]{Polianski}%
  \BibitemOpen
  \bibfield  {author} {\bibinfo {author} {\bibfnamefont {M.~L.}\ \bibnamefont
  {Polianski}}\ and\ \bibinfo {author} {\bibfnamefont {P.~W.}\ \bibnamefont
  {Brouwer}},\ }\href {\doibase 10.1103/physrevlett.92.026602} {\bibfield
  {journal} {\bibinfo  {journal} {Physical Review Letters}\ }\textbf {\bibinfo
  {volume} {92}} (\bibinfo {year} {2004}),\
  10.1103/physrevlett.92.026602}\BibitemShut {NoStop}%
\bibitem [{\citenamefont {Jungwirth}\ \emph {et~al.}(2016)\citenamefont
  {Jungwirth}, \citenamefont {Marti}, \citenamefont {Wadley},\ and\
  \citenamefont {Wunderlich}}]{Jungwirth2016}%
  \BibitemOpen
  \bibfield  {author} {\bibinfo {author} {\bibfnamefont {T.}~\bibnamefont
  {Jungwirth}}, \bibinfo {author} {\bibfnamefont {X.}~\bibnamefont {Marti}},
  \bibinfo {author} {\bibfnamefont {P.}~\bibnamefont {Wadley}}, \ and\ \bibinfo
  {author} {\bibfnamefont {J.}~\bibnamefont {Wunderlich}},\ }\href {\doibase
  10.1038/nnano.2016.18} {\bibfield  {journal} {\bibinfo  {journal} {Nature
  Nanotechnology}\ }\textbf {\bibinfo {volume} {11}},\ \bibinfo {pages} {231}
  (\bibinfo {year} {2016})}\BibitemShut {NoStop}%
\bibitem [{\citenamefont {Ando}\ \emph {et~al.}(2008)\citenamefont {Ando},
  \citenamefont {Takahashi}, \citenamefont {Harii}, \citenamefont {Sasage},
  \citenamefont {Ieda}, \citenamefont {Maekawa},\ and\ \citenamefont
  {Saitoh}}]{Ando2008}%
  \BibitemOpen
  \bibfield  {author} {\bibinfo {author} {\bibfnamefont {K.}~\bibnamefont
  {Ando}}, \bibinfo {author} {\bibfnamefont {S.}~\bibnamefont {Takahashi}},
  \bibinfo {author} {\bibfnamefont {K.}~\bibnamefont {Harii}}, \bibinfo
  {author} {\bibfnamefont {K.}~\bibnamefont {Sasage}}, \bibinfo {author}
  {\bibfnamefont {J.}~\bibnamefont {Ieda}}, \bibinfo {author} {\bibfnamefont
  {S.}~\bibnamefont {Maekawa}}, \ and\ \bibinfo {author} {\bibfnamefont
  {E.}~\bibnamefont {Saitoh}},\ }\href {\doibase
  10.1103/physrevlett.101.036601} {\bibfield  {journal} {\bibinfo  {journal}
  {Physical Review Letters}\ }\textbf {\bibinfo {volume} {101}} (\bibinfo
  {year} {2008}),\ 10.1103/physrevlett.101.036601}\BibitemShut {NoStop}%
\bibitem [{\citenamefont {Liu}\ \emph {et~al.}(2012)\citenamefont {Liu},
  \citenamefont {Pai}, \citenamefont {Li}, \citenamefont {Tseng}, \citenamefont
  {Ralph},\ and\ \citenamefont {Buhrman}}]{Liu555}%
  \BibitemOpen
  \bibfield  {author} {\bibinfo {author} {\bibfnamefont {L.}~\bibnamefont
  {Liu}}, \bibinfo {author} {\bibfnamefont {C.-F.}\ \bibnamefont {Pai}},
  \bibinfo {author} {\bibfnamefont {Y.}~\bibnamefont {Li}}, \bibinfo {author}
  {\bibfnamefont {H.~W.}\ \bibnamefont {Tseng}}, \bibinfo {author}
  {\bibfnamefont {D.~C.}\ \bibnamefont {Ralph}}, \ and\ \bibinfo {author}
  {\bibfnamefont {R.~A.}\ \bibnamefont {Buhrman}},\ }\href {\doibase
  10.1126/science.1218197} {\bibfield  {journal} {\bibinfo  {journal}
  {Science}\ }\textbf {\bibinfo {volume} {336}},\ \bibinfo {pages} {555}
  (\bibinfo {year} {2012})},\ \Eprint
  {http://arxiv.org/abs/http://science.sciencemag.org/content/336/6081/555.full.pdf}
  {http://science.sciencemag.org/content/336/6081/555.full.pdf} \BibitemShut
  {NoStop}%
\bibitem [{\citenamefont {Nakayama}\ \emph {et~al.}(2013)\citenamefont
  {Nakayama}, \citenamefont {Althammer}, \citenamefont {Chen}, \citenamefont
  {Uchida}, \citenamefont {Kajiwara}, \citenamefont {Kikuchi}, \citenamefont
  {Ohtani}, \citenamefont {Gepr\"{a}gs}, \citenamefont {Opel}, \citenamefont
  {Takahashi}, \citenamefont {Gross}, \citenamefont {Bauer}, \citenamefont
  {Goennenwein},\ and\ \citenamefont {Saitoh}}]{Nakayama2013}%
  \BibitemOpen
  \bibfield  {author} {\bibinfo {author} {\bibfnamefont {H.}~\bibnamefont
  {Nakayama}}, \bibinfo {author} {\bibfnamefont {M.}~\bibnamefont {Althammer}},
  \bibinfo {author} {\bibfnamefont {Y.-T.}\ \bibnamefont {Chen}}, \bibinfo
  {author} {\bibfnamefont {K.}~\bibnamefont {Uchida}}, \bibinfo {author}
  {\bibfnamefont {Y.}~\bibnamefont {Kajiwara}}, \bibinfo {author}
  {\bibfnamefont {D.}~\bibnamefont {Kikuchi}}, \bibinfo {author} {\bibfnamefont
  {T.}~\bibnamefont {Ohtani}}, \bibinfo {author} {\bibfnamefont
  {S.}~\bibnamefont {Gepr\"{a}gs}}, \bibinfo {author} {\bibfnamefont
  {M.}~\bibnamefont {Opel}}, \bibinfo {author} {\bibfnamefont {S.}~\bibnamefont
  {Takahashi}}, \bibinfo {author} {\bibfnamefont {R.}~\bibnamefont {Gross}},
  \bibinfo {author} {\bibfnamefont {G.~E.~W.}\ \bibnamefont {Bauer}}, \bibinfo
  {author} {\bibfnamefont {S.~T.~B.}\ \bibnamefont {Goennenwein}}, \ and\
  \bibinfo {author} {\bibfnamefont {E.}~\bibnamefont {Saitoh}},\ }\href
  {\doibase 10.1103/physrevlett.110.206601} {\bibfield  {journal} {\bibinfo
  {journal} {Physical Review Letters}\ }\textbf {\bibinfo {volume} {110}}
  (\bibinfo {year} {2013}),\ 10.1103/physrevlett.110.206601}\BibitemShut
  {NoStop}%
\bibitem [{\citenamefont {Turgut}\ \emph {et~al.}(2016)\citenamefont {Turgut},
  \citenamefont {Zusin}, \citenamefont {Legut}, \citenamefont {Carva},
  \citenamefont {Knut}, \citenamefont {Shaw}, \citenamefont {Chen},
  \citenamefont {Tao}, \citenamefont {Nembach}, \citenamefont {Silva},
  \citenamefont {Mathias}, \citenamefont {Aeschlimann}, \citenamefont
  {Oppeneer}, \citenamefont {Kapteyn}, \citenamefont {Murnane},\ and\
  \citenamefont {Grychtol}}]{Turgut2016}%
  \BibitemOpen
  \bibfield  {author} {\bibinfo {author} {\bibfnamefont {E.}~\bibnamefont
  {Turgut}}, \bibinfo {author} {\bibfnamefont {D.}~\bibnamefont {Zusin}},
  \bibinfo {author} {\bibfnamefont {D.}~\bibnamefont {Legut}}, \bibinfo
  {author} {\bibfnamefont {K.}~\bibnamefont {Carva}}, \bibinfo {author}
  {\bibfnamefont {R.}~\bibnamefont {Knut}}, \bibinfo {author} {\bibfnamefont
  {J.~M.}\ \bibnamefont {Shaw}}, \bibinfo {author} {\bibfnamefont
  {C.}~\bibnamefont {Chen}}, \bibinfo {author} {\bibfnamefont {Z.}~\bibnamefont
  {Tao}}, \bibinfo {author} {\bibfnamefont {H.~T.}\ \bibnamefont {Nembach}},
  \bibinfo {author} {\bibfnamefont {T.~J.}\ \bibnamefont {Silva}}, \bibinfo
  {author} {\bibfnamefont {S.}~\bibnamefont {Mathias}}, \bibinfo {author}
  {\bibfnamefont {M.}~\bibnamefont {Aeschlimann}}, \bibinfo {author}
  {\bibfnamefont {P.~M.}\ \bibnamefont {Oppeneer}}, \bibinfo {author}
  {\bibfnamefont {H.~C.}\ \bibnamefont {Kapteyn}}, \bibinfo {author}
  {\bibfnamefont {M.~M.}\ \bibnamefont {Murnane}}, \ and\ \bibinfo {author}
  {\bibfnamefont {P.}~\bibnamefont {Grychtol}},\ }\href {\doibase
  10.1103/PhysRevB.94.220408} {\bibfield  {journal} {\bibinfo  {journal} {Phys.
  Rev. B}\ }\textbf {\bibinfo {volume} {94}},\ \bibinfo {pages} {220408}
  (\bibinfo {year} {2016})}\BibitemShut {NoStop}%
\bibitem [{\citenamefont {Razdolski}\ \emph {et~al.}(2017)\citenamefont
  {Razdolski}, \citenamefont {Alekhin}, \citenamefont {Ilin}, \citenamefont
  {Meyburg}, \citenamefont {Roddatis}, \citenamefont {Diesing}, \citenamefont
  {Bovensiepen},\ and\ \citenamefont {Melnikov}}]{NatComm82017}%
  \BibitemOpen
  \bibfield  {author} {\bibinfo {author} {\bibfnamefont {I.}~\bibnamefont
  {Razdolski}}, \bibinfo {author} {\bibfnamefont {A.}~\bibnamefont {Alekhin}},
  \bibinfo {author} {\bibfnamefont {N.}~\bibnamefont {Ilin}}, \bibinfo {author}
  {\bibfnamefont {J.~P.}\ \bibnamefont {Meyburg}}, \bibinfo {author}
  {\bibfnamefont {V.}~\bibnamefont {Roddatis}}, \bibinfo {author}
  {\bibfnamefont {D.}~\bibnamefont {Diesing}}, \bibinfo {author} {\bibfnamefont
  {U.}~\bibnamefont {Bovensiepen}}, \ and\ \bibinfo {author} {\bibfnamefont
  {A.}~\bibnamefont {Melnikov}},\ }\href {\doibase 10.1038/ncomms15007}
  {\bibfield  {journal} {\bibinfo  {journal} {Nature Communications}\ }\textbf
  {\bibinfo {volume} {8}} (\bibinfo {year} {2017}),\
  10.1038/ncomms15007}\BibitemShut {NoStop}%
\bibitem [{\citenamefont {Rudolf}\ \emph {et~al.}(2012)\citenamefont {Rudolf},
  \citenamefont {La-O-Vorakiat}, \citenamefont {Battiato}, \citenamefont
  {Adam}, \citenamefont {Shaw}, \citenamefont {Turgut}, \citenamefont
  {Maldonado}, \citenamefont {Mathias}, \citenamefont {Grychtol}, \citenamefont
  {Nembach}, \citenamefont {Silva}, \citenamefont {Aeschlimann}, \citenamefont
  {Kapteyn}, \citenamefont {Murnane}, \citenamefont {Schneider},\ and\
  \citenamefont {Oppeneer}}]{Rudolf2012}%
  \BibitemOpen
  \bibfield  {author} {\bibinfo {author} {\bibfnamefont {D.}~\bibnamefont
  {Rudolf}}, \bibinfo {author} {\bibfnamefont {C.}~\bibnamefont
  {La-O-Vorakiat}}, \bibinfo {author} {\bibfnamefont {M.}~\bibnamefont
  {Battiato}}, \bibinfo {author} {\bibfnamefont {R.}~\bibnamefont {Adam}},
  \bibinfo {author} {\bibfnamefont {J.~M.}\ \bibnamefont {Shaw}}, \bibinfo
  {author} {\bibfnamefont {E.}~\bibnamefont {Turgut}}, \bibinfo {author}
  {\bibfnamefont {P.}~\bibnamefont {Maldonado}}, \bibinfo {author}
  {\bibfnamefont {S.}~\bibnamefont {Mathias}}, \bibinfo {author} {\bibfnamefont
  {P.}~\bibnamefont {Grychtol}}, \bibinfo {author} {\bibfnamefont {H.~T.}\
  \bibnamefont {Nembach}}, \bibinfo {author} {\bibfnamefont {T.~J.}\
  \bibnamefont {Silva}}, \bibinfo {author} {\bibfnamefont {M.}~\bibnamefont
  {Aeschlimann}}, \bibinfo {author} {\bibfnamefont {H.~C.}\ \bibnamefont
  {Kapteyn}}, \bibinfo {author} {\bibfnamefont {M.~M.}\ \bibnamefont
  {Murnane}}, \bibinfo {author} {\bibfnamefont {C.~M.}\ \bibnamefont
  {Schneider}}, \ and\ \bibinfo {author} {\bibfnamefont {P.~M.}\ \bibnamefont
  {Oppeneer}},\ }\href {\doibase 10.1038/ncomms2029} {\bibfield  {journal}
  {\bibinfo  {journal} {Nature Communications}\ }\textbf {\bibinfo {volume}
  {3}} (\bibinfo {year} {2012}),\ 10.1038/ncomms2029}\BibitemShut {NoStop}%
\bibitem [{\citenamefont {Hatami}\ \emph {et~al.}(2007)\citenamefont {Hatami},
  \citenamefont {Bauer}, \citenamefont {Zhang},\ and\ \citenamefont
  {Kelly}}]{PhysRevLett.99.066603}%
  \BibitemOpen
  \bibfield  {author} {\bibinfo {author} {\bibfnamefont {M.}~\bibnamefont
  {Hatami}}, \bibinfo {author} {\bibfnamefont {G.~E.~W.}\ \bibnamefont
  {Bauer}}, \bibinfo {author} {\bibfnamefont {Q.}~\bibnamefont {Zhang}}, \ and\
  \bibinfo {author} {\bibfnamefont {P.~J.}\ \bibnamefont {Kelly}},\ }\href
  {\doibase 10.1103/PhysRevLett.99.066603} {\bibfield  {journal} {\bibinfo
  {journal} {Phys. Rev. Lett.}\ }\textbf {\bibinfo {volume} {99}},\ \bibinfo
  {pages} {066603} (\bibinfo {year} {2007})}\BibitemShut {NoStop}%
\bibitem [{\citenamefont {Foros}\ \emph {et~al.}(2005)\citenamefont {Foros},
  \citenamefont {Brataas}, \citenamefont {Tserkovnyak},\ and\ \citenamefont
  {Bauer}}]{PhysRevLett.95.016601}%
  \BibitemOpen
  \bibfield  {author} {\bibinfo {author} {\bibfnamefont {J.}~\bibnamefont
  {Foros}}, \bibinfo {author} {\bibfnamefont {A.}~\bibnamefont {Brataas}},
  \bibinfo {author} {\bibfnamefont {Y.}~\bibnamefont {Tserkovnyak}}, \ and\
  \bibinfo {author} {\bibfnamefont {G.~E.~W.}\ \bibnamefont {Bauer}},\ }\href
  {\doibase 10.1103/PhysRevLett.95.016601} {\bibfield  {journal} {\bibinfo
  {journal} {Phys. Rev. Lett.}\ }\textbf {\bibinfo {volume} {95}},\ \bibinfo
  {pages} {016601} (\bibinfo {year} {2005})}\BibitemShut {NoStop}%
\bibitem [{\citenamefont {Wegrowe}(2010)}]{Wegrowe2010}%
  \BibitemOpen
  \bibfield  {author} {\bibinfo {author} {\bibfnamefont {J.-E.}\ \bibnamefont
  {Wegrowe}},\ }\href {\doibase 10.1016/j.ssc.2009.10.046} {\bibfield
  {journal} {\bibinfo  {journal} {Solid State Communications}\ }\textbf
  {\bibinfo {volume} {150}},\ \bibinfo {pages} {519} (\bibinfo {year}
  {2010})}\BibitemShut {NoStop}%
\bibitem [{\citenamefont {Zhang}\ \emph {et~al.}(2016)\citenamefont {Zhang},
  \citenamefont {Bai}, \citenamefont {Chen}, \citenamefont {Guo}, \citenamefont
  {Fan}, \citenamefont {Xue}, \citenamefont {Houssameddine},\ and\
  \citenamefont {Hu}}]{PhysRevB.94.064414}%
  \BibitemOpen
  \bibfield  {author} {\bibinfo {author} {\bibfnamefont {Z.}~\bibnamefont
  {Zhang}}, \bibinfo {author} {\bibfnamefont {L.}~\bibnamefont {Bai}}, \bibinfo
  {author} {\bibfnamefont {X.}~\bibnamefont {Chen}}, \bibinfo {author}
  {\bibfnamefont {H.}~\bibnamefont {Guo}}, \bibinfo {author} {\bibfnamefont
  {X.~L.}\ \bibnamefont {Fan}}, \bibinfo {author} {\bibfnamefont {D.~S.}\
  \bibnamefont {Xue}}, \bibinfo {author} {\bibfnamefont {D.}~\bibnamefont
  {Houssameddine}}, \ and\ \bibinfo {author} {\bibfnamefont {C.-M.}\
  \bibnamefont {Hu}},\ }\href {\doibase 10.1103/PhysRevB.94.064414} {\bibfield
  {journal} {\bibinfo  {journal} {Phys. Rev. B}\ }\textbf {\bibinfo {volume}
  {94}},\ \bibinfo {pages} {064414} (\bibinfo {year} {2016})}\BibitemShut
  {NoStop}%
\bibitem [{\citenamefont {Choi}\ \emph {et~al.}(2015)\citenamefont {Choi},
  \citenamefont {Moon}, \citenamefont {Min}, \citenamefont {Lee},\ and\
  \citenamefont {Cahill}}]{Choi2015}%
  \BibitemOpen
  \bibfield  {author} {\bibinfo {author} {\bibfnamefont {G.-M.}\ \bibnamefont
  {Choi}}, \bibinfo {author} {\bibfnamefont {C.-H.}\ \bibnamefont {Moon}},
  \bibinfo {author} {\bibfnamefont {B.-C.}\ \bibnamefont {Min}}, \bibinfo
  {author} {\bibfnamefont {K.-J.}\ \bibnamefont {Lee}}, \ and\ \bibinfo
  {author} {\bibfnamefont {D.~G.}\ \bibnamefont {Cahill}},\ }\href {\doibase
  10.1038/nphys3355} {\bibfield  {journal} {\bibinfo  {journal} {Nature
  Physics}\ }\textbf {\bibinfo {volume} {11}} (\bibinfo {year} {2015}),\
  10.1038/nphys3355}\BibitemShut {NoStop}%
\bibitem [{\citenamefont {Belashchenko}\ \emph {et~al.}(2016)\citenamefont
  {Belashchenko}, \citenamefont {Kovalev},\ and\ \citenamefont {van
  Schilfgaarde}}]{Belashchenko}%
  \BibitemOpen
  \bibfield  {author} {\bibinfo {author} {\bibfnamefont {K.~D.}\ \bibnamefont
  {Belashchenko}}, \bibinfo {author} {\bibfnamefont {A.~A.}\ \bibnamefont
  {Kovalev}}, \ and\ \bibinfo {author} {\bibfnamefont {M.}~\bibnamefont {van
  Schilfgaarde}},\ }\href {\doibase 10.1103/PhysRevLett.117.207204} {\bibfield
  {journal} {\bibinfo  {journal} {Phys. Rev. Lett.}\ }\textbf {\bibinfo
  {volume} {117}},\ \bibinfo {pages} {207204} (\bibinfo {year}
  {2016})}\BibitemShut {NoStop}%
\end{thebibliography}%
\bibliographystyle{apsrev4-1}

\end{document}